# Parametric Modelling Within Immersive Environments

*Building a Bridge Between Existing Tools and Virtual Reality Headsets*

Adrien Coppens[1], Tom Mens[2], Mohamed-Anis Gallas[3]
[1,2]Software Engineering Lab, University of Mons [3]Architectural Design Unit, University of Mons
[1,2,3]{adrien.coppens|tom.mens|mohamed-anis}@umons.ac.be

*Even though architectural modelling radically evolved over the course of its history, the current integration of Augmented Reality (AR) and Virtual Reality (VR) components in the corresponding design tasks is mostly limited to enhancing visualisation. Little to none of these tools attempt to tackle the challenge of modelling within immersive environments, that calls for new input modalities in order to move away from the traditional mouse and keyboard combination. In fact, relying on 2D devices for 3D manipulations does not seem to be effective as it does not offer the same degrees of freedom. We therefore present a solution that brings VR modelling capabilities to Grasshopper, a popular parametric design tool. Together with its associated proof-of-concept application, our extension offers a glimpse at new perspectives in that field. By taking advantage of them, one can edit geometries with real-time feedback on the generated models, without ever leaving the virtual environment. The distinctive characteristics of VR applications provide a range of benefits without obstructing design activities. The designer can indeed experience the architectural models at full scale from a realistic point-of-view and truly feels immersed right next to them.*

## INTRODUCTION

The AEC (Architecture, Engineering and Construction) industry has radically evolved due to technological improvements that gave rise to advanced automated tools for the corresponding design tasks. Traditional paper drawings and scale models are now accompanied by 2D and 3D modelling software, but also by real-time (lighting, structural) simulations and photo-realistic renderings of architectural designs. While these tools are constantly improving in terms of performance and accuracy, their modalities of user interaction do not follow the same trend.

Experiences combining virtual and real elements, labelled as Mixed Reality (MR), can be placed on Milgram's Reality-Virtuality continuum (Milgram and Kishino 1994). An extended version of that continuum is depicted in Figure 1 and encompasses experiences that range from "close to reality" to "almost entirely virtual". Augmented Reality (superimposing virtual elements on real world objects) and Virtual Reality (complete immersion in a virtual environment) elements have started to appear in CAAD. However, their purpose is mostly limited to enhancing the visualisation of models designed using desktop software.

Humans are accustomed to manipulating 3D objects intuitively in the real world as the corresponding interactions are also 3-dimensional. However, virtual models designed with computer software are typically harder to deal with because most tools still rely on a traditional mouse and keyboard combination to express user intent. The current state of the practice in CAAD is no different: users need to work around the limitations imposed by lower-dimensional interaction devices. For example, to navigate through 3D models, one often needs to switch between "modes" (e.g. translation and rotation) that may be further subdivided in 3 dimensions, whereas 3D navigation can happen much more seamlessly in the real or VR world. As a concrete illustration, rotating objects in the Rhino 3D modeller generally involves the use of a rotation tool that only controls one axis at a time. To rotate a model around all 3 "standard" axes (traditional x, y and z axes), one typically performs three separate actions, which means that it is necessary to anticipate the effect of those rotations. There is also a special tool that allows users to define an arbitrary axis and rotate around it. Even though this feature permits single rotations that involve all 3 standard axes, there still is only one degree of freedom (you cannot rotate around another axis at the same time) and defining that axis according to the foreseen rotation is generally a difficult task.

The above shortcomings, as well as potential solutions, have been pointed out for over two decades (Chu et al. 1997) but these solutions have not been integrated yet in contemporary CAAD software. As a result, very few tools tackle the challenge of architectural modelling within 3D immersive environments, as this requires new input modalities that go beyond the traditional interaction paradigms.

Parametric modelling already transforms the design process by relying on visual algorithms to define models. As a first step towards bringing that approach in VR, we built an extension for Grasshopper, a parametric design plug-in for Rhino. This extension allows users to see and modify 3D geometries using an HTC Vive VR headset, to benefit from real-time interaction with, and feedback on, the generated models.

## RELATED WORK

Several tools attempted to provide annotation and sketching capabilities for architectural models in VR. IrisVR Prospect provides compatibility with many popular tools. Similarly, Twinmotion includes the ability to create VR visualisations from models designed with the system. Hyve-3D (Dorta et al. 2016) suggests the use of a tablet, controlling a 3D cursor to move around and enable sketching.

Integrating VR solutions with traditional CAAD tools necessitates an automatic synchronisation between that tool's model and its VR counterpart. Forcing the user to alternate between the desktop application and the VR tool through an edit-export-validate procedure for every modification is obtrusive to the design process.

Previous work [1] proposed a one-way synchronisation from Grasshopper to the Unity game engine. Using this solution, one can create VR applications to visualise geometries generated from Grasshopper algorithms, with automatic updates when the model changes. However, at this point, the VR application only receives meshes (i.e. simplified 3D models used for graphical rendering). No semantic information is exchanged about the underlying algorithm and the only way to edit the model is within Grasshopper.

As far as we know, no previous contribution provides parametric modelling capabilities within VR and that field therefore has yet to be explored.

## PROPOSED SOLUTION

Our first effort in filling that void involved the development of three software components. The first one is a plug-in that connects Grasshopper to a Web-

Figure 1
Milgram et al's RV continuum encompassing mixed reality experiences

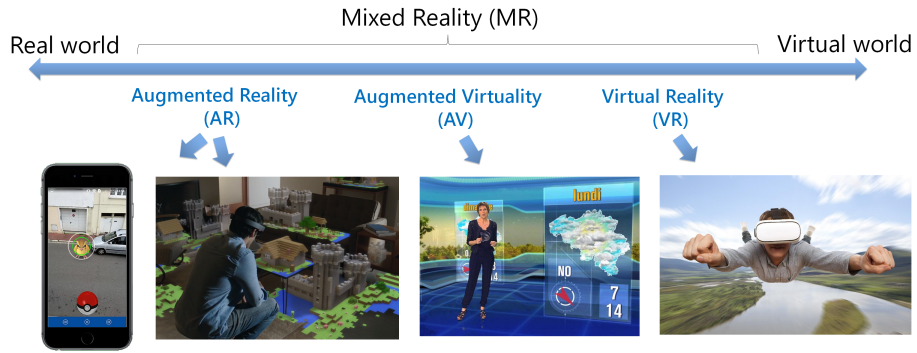

Socket server in order to enable communication with client applications. The second component is the web server, which acts as a relay between the plug-in and the client application. The third component is the client application itself. We developed a proof-of-concept of such an application running on the HTC Vive VR headset.

The Grasshopper plug-in converts meshes to a binary format and sends them to the server for each change. It also announces what parameters can be modified by the client application (including its type, the corresponding restrictions and its current value). Once the client application receives that information, the given models can be rendered and their corresponding parameter controls (e.g. sliders) can be displayed and interacted with. As some Grasshopper parameters have an excessive amount of possible values for a proper interaction in VR, we need to reduce the number of available steps in the VR interface (we chose to set the limit at approximately 20 fairly distributed steps). Modifications made to those values are sent back to the Grasshopper plug-in that replicates them on the actual model parameters. A video [2] accompanying this paper demonstrates the use of our proof-of-concept application while Figure 2 provides a general overview.

The interconnections between those components allow the designer to modify values and observe the resulting model changes in real-time, taking advantage of VR characteristics without having to endure a more complex and time-consuming workflow of juggling with the VR application and Grasshopper. VR enables to move the architectural modelling process another step closer to user-centred design, involving end-users earlier in the process. Indeed, it is much easier for non-experts to visualise and understand a designed element based on a 3D virtual model as opposed to its multiple 2D projections. The proposed tool therefore enables a much easier collaboration that allows to fine-tune designed models with direct feedback from the client.

## SWOT ANALYSIS

A SWOT (Strengths, Weaknesses, Opportunities, Threats) analysis is a technique often used to evaluate business strategies. It studies the Strengths and Weaknesses of an approach as well as the Opportunities and Threats coming from its surrounding environment. For this reason, conducting such an analysis helps in assessing the relevance and the consistency of a specific strategic decision. In our case, we use a SWOT analysis to evaluate how the use of VR compares with traditional CAAD tools as well as to identify the benefits our proposed solution can provide and the concerns it may raise.

The outcome of that analysis is described in the remaining parts of this section, while Table 1 presents a summary overview.

| Strengths: | Weaknesses: |
|---|---|
| • Sense of presence<br>• Perspective view and notion of scale<br>• Simplified feedback loop (live interaction) | • Reduced accuracy<br>• Discomfort for long exposures |
| Opportunities: | Threats: |
| • Intuitive manipulation<br>• User-centred design | • Resistance to change<br>• (Small) additional cost |

Table 1
Summarised SWOT analysis for the proposed solution

## Strengths

VR enhances our ability to visualise conceptual models with a set of interesting characteristics. Being able to see full-scale models from a realistic perspective helps users in having a better idea of what the final products will look like. The sense of presence also makes it possible to experience being next to or even inside that model.

Existing VR integrations generally require a set of actions from the user to export a model to the virtual environment. Once problems are detected or enhancements are suggested, the designer must return to the desktop software to carry out the corresponding modifications. He can only visualise them in VR by going through the export process once again. Our solution simplifies that feedback loop through a real-time synchronisation between the desktop software and the VR application. This minimises the overhead on the design process while enabling live interactions with the user.

## Threats

Despite the expected benefits, some factors might limit the adoption of VR in CAAD solutions. Current threats include the additional cost of the hardware; as well as possible resistance to change by experienced designers that are accustomed to traditional desktop based tools. Even though compelling VR systems have become affordable, potential users may remain sceptical due to past experiences with less optimal devices.

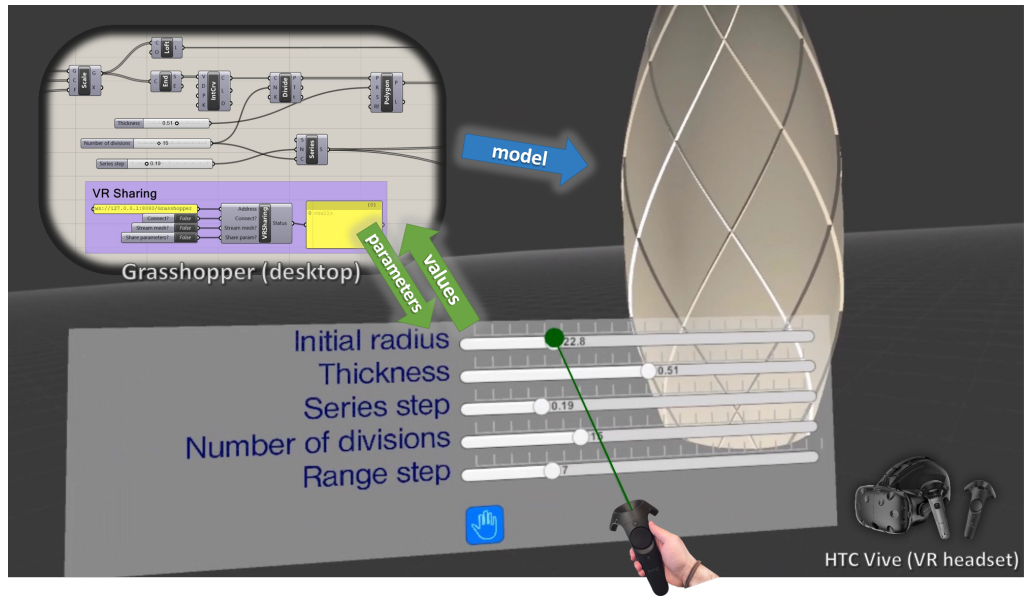

Figure 2
Overview of our proof-of-concept VR application

## Opportunities

On the input level, the mismatch between 3D manipulations and traditional 2D devices could be solved by relying on gestural interaction. This technique often accompanies VR headsets thanks to 6-DoF (degrees of freedom) trackers that can capture 3D translations and rotations. This allows for a much more natural and intuitive manipulation of 3D models and enables simultaneous positional and rotational modifications. Table 2 compares the standard computer mouse to the Vive 6-DoF controller in terms of degrees of freedom to point out how adapted they seem to be to perform 2D and 3D transformations.

Table 2
Available or required degrees of freedom depending on the device or transformation type

|  | Degrees of Freedom (DoF) | | | | | |
|---|---|---|---|---|---|---|
|  | Positional | | | Rotational | | |
|  | x | y | z | x | y | z |
| Standard mouse | ● | ● |  |  |  |  |
| Vive controller | ● | ● | ○ | ○ | ○ | ○ |
| 2D transformation | ● | ● |  |  |  | ○ |
| 3D transformation | ● | ● | ○ | ○ | ○ | ○ |

● present for all items   ○ missing on some items

On another note, a VR visualisation of an architectural model is much easier to understand than a traditional plan from a layman point of view. This means clients can be involved in the design process by giving feedback much earlier, which would enable a User-Centred Design (UCD) approach.

## Weaknesses

Aside from a potential discomfort for long exposures, the benefits in terms of easier manipulation and intuitiveness are generally assumed to go along with a reduced accuracy when performing such movements. (Bérard et al. 2009) tend to concur with that statement; while others (McMahan et al. 2006; Hinckley et al. 1997) suggest that devices with more degrees of freedom are more appropriate than a standard mouse for 3-dimensional transformations.

Since those publications, significant progress in terms of hardware has been achieved, and new solutions focused on VR for CA(A)D solutions have emerged. Hence it may be interesting to carry out new usability studies in such modern settings.

# FUTURE WORK

In its current state, our proposed solution only allows to modify parameter values and assess the resulting effects. This already opens new perspectives for parametric modelling but we aim to push the concept even further by allowing designers to modify the underlying 6algorithm that describes the parametric model (and that is interpreted by Grasshopper to produce the geometry), without ever leaving the virtual environment. Figure 3 compares our VR application to the standard Rhino 3D + Grasshopper combination in terms of which kind of information is available about the designed model. While our solution permits parameter sharing (it will later allow to edit the algorithm as well; hence the dashed lines on the figure), only the rendering-ready (mesh) version of the model is shared. There is no plan on integrating any kind of intermediary "semantic" model such as a boundary representation into the VR application. Generating the designed model from the algorithm will continue to be performed by Grasshopper.

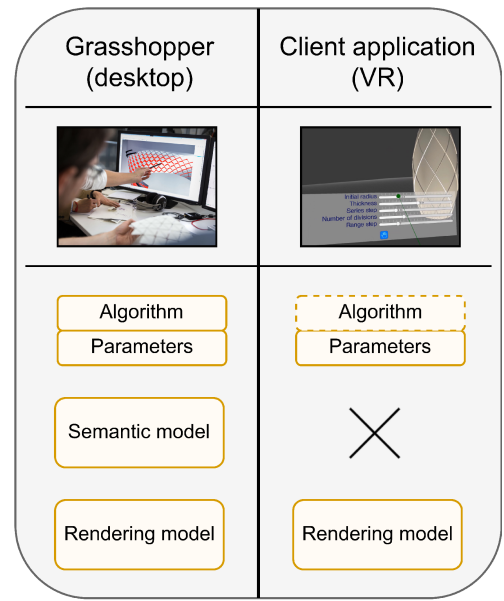

Figure 3
Comparing Grasshopper (with Rhino 3D) and our VR application in terms of available information

As for the Human-Computer Interaction aspect, the user interface of our current proof-of-concept is limited since it is represented using object-space WIMP (Windows, Icons, Menus and Pointing device): it basically positions flat screen-like controls in the virtual 3D space. We aim to integrate more advanced interaction paradigms based on post-WIMP interfaces, relying on a combination of gestures and voice commands.

Such interfaces will allow to extend the VR functionality beyond parameter sharing, with features enabling zooming and scaling as well as component selection and editing.

On a shorter notice, we plan on adding collaborative capabilities to our solution as design processes often involve multiple participants. Adding such features means that several challenges will need to be addressed such as conflicting intents and connectivity/latency issues. Furthermore, the impact of those issues (and therefore the importance of properly managing them) strongly increases when it comes to remote (distant) collaboration.

## CONCLUSION

We presented a proof-of-concept prototype for parametric modelling within immersive environments by building a bridge between Grasshopper and a VR headset. We compared it with existing work and considered the potential benefits and the associated drawbacks, by carrying out a SWOT analysis. Preliminary tests, with users having a few years of experience in architectural design, has collected promising feedback on the present set of features.

But, as mentioned in the "future work" section, further development is to be expected in terms of functionality and usability. It seems necessary to go beyond traditional types of interfaces and the urge to embrace new interaction paradigms is indeed strengthened by the tendency of desktop-based CAD interfaces to quickly become very complex.